\documentclass[a4paper,cleveref, autoref, thm-restate, table]{article}

\usepackage{amsthm}
\usepackage{amssymb}
\usepackage{graphicx} 
\usepackage{mathtools}
\usepackage{varwidth}
\usepackage{authblk}

\usepackage{todonotes}
\usepackage{hyperref}
\usepackage{comment}

\usepackage{algorithm}
\usepackage[noend]{algpseudocode}
\usepackage{xspace}
\usepackage[short,nocomma]{optidef}
\usepackage{makecell}
\algrenewcommand\algorithmicindent{1.0em}
\usepackage{thmtools}
\declaretheorem[name=Theorem]{thm}
\declaretheorem[name=Corollary]{cor}

\usepackage{amsmath}
\usepackage{mathabx}
\usepackage{tikz}
\usetikzlibrary{positioning}
\usetikzlibrary{shapes}
\usetikzlibrary{calc}
\usetikzlibrary{decorations.pathreplacing}
\usetikzlibrary{matrix}

\usepackage{cleveref}

\usepackage{nicematrix}

\usepackage{diagbox}
\usepackage{hhline}

\definecolor{mygreen}{RGB}{0, 102, 0}
\definecolor{myred}{RGB}{153, 0, 0}

\newcommand{\mO}{\mathcal{O}}
\newcommand{\conc}{\ensuremath{\circ}}
\newcommand{\realph}[1]{\ensuremath{\Sigma_{#1}}\xspace}

\theoremstyle{definition}
\newtheorem*{conjecture*}{Conjecture}
\newtheorem{remark}{Remark}
\newtheorem{observation}{Observation}
\newtheorem{definition}{Definition}
\newtheorem{example}{Example}

\theoremstyle{plain}
\newtheorem{lemma}{Lemma}

\newcommand\ov{\textsf{OV}\xspace}

\newcommand\uovh{\textsf{UOVH}\xspace}

\newcommand\seth{\textsf{SETH}\xspace}

\newcommand{\cA}{\mathcal{A}}
\newcommand{\cB}{\mathcal{B}}
\newcommand{\cL}{\mathcal{L}}
\newcommand{\cX}{\mathcal{X}}
\newcommand{\cY}{\mathcal{Y}}

\def\dd{\mathinner{.\,.}}

\DeclareMathOperator{\lcm}{lcm}

\title{Are Depth-2 Regular Expressions\\ Hard to Intersect?}

\author[1]{Rocco Ascone}
\author[2]{Giulia Bernardini}
\author[3]{Alessio Conte}
\author[3]{Veronica Guerrini}
\author[3]{Giulia Punzi}

\affil[1]{University of Trieste, Italy\\
\texttt{rocco.ascone@phd.units.it}}
\affil[2]{University of Milan, Italy\\
\texttt{giulia.bernardini@unimi.it}}
\affil[3]{University of Pisa, Italy\\
\texttt{\{alessio.conte, veronica.guerrini, giulia.punzi\}@unipi.it}}

\date{}

\begin{document}

\maketitle

\begin{abstract}

    We study the basic regular expression intersection testing problem, which asks to determine whether the intersection of the languages of two regular expressions is nonempty.
    A textbook solution to this problem is to construct the nondeterministic finite automaton that accepts the language of both expressions. 
    This procedure results in a $\Theta(mn)$ running time, where $m$ and $n$ are the sizes of the two expressions, respectively.
    Following the approach of Backurs and Indyk [FOCS'16] and Bringmann, Grønlund, and Larsen [FOCS'17] on regular expression matching and membership testing, we study the complexity of intersection testing for homogeneous regular expressions of bounded depth involving concatenation, OR, Kleene star, and Kleene plus. Specifically, we consider all combinations of types of depth-2 regular expressions and classify the time complexity of intersection testing as either linear or quadratic, assuming SETH.
    The most interesting result is a quadratic conditional lower bound for testing the intersection of a ``concatenation of +s'' expression with a ``concatenation of ORs'' expression: this is the only hard case that does not involve the Kleene star operator and is not implied by existing lower bounds for the simpler membership testing problem.
\end{abstract}

\newpage 
\setcounter{page}{1}

\section{Introduction}\label{sec:intro}

A regular expression (regexp) $\cA$ over an alphabet $\Sigma$ is a formula that encodes a set of strings over $\Sigma$, called the \emph{language} of $\cA$, using the operators concatenation ($\conc$), OR ($|$), Kleene Star ($*$) and Kleene plus ($+$), formally defined in \Cref{sec:preliminaries}. The \emph{regular expression intersection testing} problem asks, given two regexps $\cA$ and $\cB$, to determine if the intersection of their languages is nonempty, that is, to decide whether a string exists that can be generated by both expressions. 
Similar to other basic tasks like regular expression pattern matching and membership testing, the intersection testing problem has a multitude of applications, including web services~\cite{DBLP:journals/jwsr/WombacherFMN04}, data privacy~\cite{DBLP:conf/pst/GuancialeGL14}, software security~\cite{DBLP:conf/uss/LiCC0PCCC21,DBLP:conf/sp/WangZLXHLYXZLH23}, model checking~\cite{DBLP:conf/tacas/HenriksenJJKPRS95}, and satisfiability modulo theories~\cite{DBLP:conf/frocos/LiangTRTB15,DBLP:conf/pldi/StanfordVB21,DBLP:journals/pacmpl/ChenFHHHKLRW22,DBLP:conf/tacas/BarbosaBBKLMMMN22}.

Given two regular expressions $\cA$ of size $m$ and $\cB$ of size $n$, a textbook solution to intersection testing is to construct in linear time the nondeterministic finite automata accepting the language of $\cA$ and $\cB$, respectively, then to compute the automaton accepting the intersection of the two languages in quadratic time $\mO(mn)$, and decide whether its language is nonempty in time $\mO(mn)$. 
It is natural to wonder whether there exist special types of regular expressions for which intersection testing can be solved exponentially faster than in quadratic time, and if, on the contrary, for some types it can be proved that no truly subquadratic algorithm exists unless some popular hypotheses are falsified.

\begin{figure}[h]
    \centering
    \begin{tikzpicture}[level distance=10mm,level 1/.style={sibling distance=10mm},
         level 2/.style={sibling distance=6mm}]
        \node (reg1) {$\conc$}
        child {node [anchor=south]{$+$} 
            child {node [anchor=south]{$b$}}}
        child {node [anchor=south]{$b$}}
        child {node [anchor=south]{$+$}
          child {node [anchor=south]{$c$}}
        }
        child {node [anchor=south]{$+$}
          child {node [anchor=south]{$c$}}
        }
        child {node [anchor=south]{$a$}};
        \node (reg2) at ([xshift=5cm]reg1) {$\conc$}
        child {node [anchor=south]{$|$} 
            child {node [anchor=south]{$a$}}
            child {node [anchor=south]{$b$}}
            }
        child {node [anchor=south]{$b$}}
        child {node [anchor=south]{$c$}}
        child {node [anchor=south]{$|$}
          child {node [anchor=south] {$a$}}
          child {node [anchor=south]{$b$}}
          child {node [anchor=south]{$c$}}
            };
        \node at ([yshift=0.4cm]reg1) {$b^+b c^+c^+a$};
        \node at ([yshift=0.4cm]reg2) {$[a|b]bc[a|b|c]$};
    \end{tikzpicture}
    \caption{Example of a regular expression of type $\conc +$ (left) and $\conc |$ (right).}
    \label{fig:tree_example}
\end{figure}
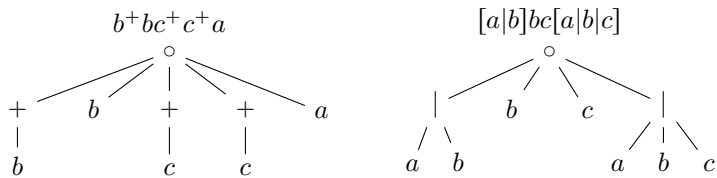

This question has been extensively considered in the literature for the related problems of regular expression membership testing and pattern matching, asking, respectively, if given a regexp $\cA$ and a string $t$, the whole $t$ or some of its fragments can be generated from $\cA$.  
Backurs and Indyk~\cite{focs16} and Bringmann, Gr\o nlund and Larsen~\cite{DBLP:conf/focs/BringmannGL17} accomplished a systematic classification of the time complexity of \emph{homogeneous} regexp pattern matching and membership testing as either quadratic or strongly subquadratic. 
A regexp is homogeneous if, given its representation as a rooted tree whose inner nodes correspond to operators and leaves correspond to letters of $\Sigma$, the inner nodes at any fixed level all correspond to the same operator. The \emph{depth} of a homogeneous regexp is the depth of this tree, and its \emph{type} is given by the sequence of operators on a root-to-leaf path of maximal length: see \Cref{fig:tree_example}.

Since a string is a depth-1 regular expression of type $\conc$, membership testing with a regexp of type $t$ can be viewed as testing the intersection for types $\conc$ and $t$. This implies that if membership testing with type $t$ is SETH-hard, so is testing the intersection of type $t$ with any other type $u$ that begins with the $\conc$ operator.
This paper systematically classifies the complexity of intersection testing for all types of homogeneous regexps of depth $\leq 2$ as either linear or quadratic, conditioned on the Strong Exponential Time Hypothesis (SETH)~\cite{DBLP:journals/jcss/ImpagliazzoP01}.

\noindent\textbf{Our results.}~
Our main results are summarised in Tables~\ref{tab:cdot_2deg_regexp} and~\ref{tab:test} and provide a full dichotomy for depth-$2$ homogeneous regexp intersection testing (blank cells are due to the symmetry of the problem).
The hardness results on the first row of \Cref{tab:cdot_2deg_regexp} are directly implied by the SETH-hardness of membership testing with regexps of type $\conc*$ proved by Backurs and Indyk~\cite{focs16}. The same authors devised near-linear-time algorithms for membership testing with all the other types of depth-$2$ homogeneous regexps. 
Interestingly, we prove that testing the intersection of a regexp of type $\conc |$ with one of type $\conc +$ is SETH-hard, 
despite the existing linear-time algorithms for membership testing with these types.
We prove this result with a highly nontrivial reduction\footnote{An alternative reduction from gapped subsequence with length constraints matching~\cite{day2022subsequences}, also conditioned on \seth, was proposed by an anonymous reviewer.} from an unbalanced version of the Orthogonal Vectors problem~\cite{10.1109/FOCS.2015.15}.
The fact that regular expressions of type $\conc +$ have a form similar to those of type $\conc *$ might lead one to think that a minor modification of the reduction proposed in~\cite{focs16} for $(\conc *)$-membership testing would suffice to prove SETH-hardness for our problem. 
However, although some of the gadgets and assumptions in our reduction are indeed similar to those of Backurs and Indyk~\cite{focs16}, their reduction relies heavily on the possibility of repeating each letter of the regexp $0$ times: so heavily that removing this possibility leads to the existence of a linear-time algorithm for $(\conc +)$-membership testing. 
The lack of this possibility in our case thus requires entirely new techniques that exploit the form of the regexp of type $\conc |$. 
Interestingly enough, it is easy to test the intersection of two regexps of the same type $\conc |$ or $\conc +$ in linear time; thus, the quadratic conditional lower bound for $(\conc +,\conc |)$-intersection testing stems from the specific interplay of the two different types.

In addition to the cases mentioned above, we show that intersection testing is solvable in linear time for all the other combinations of depth-$2$ homogeneous regexps. The algorithms we propose are simple combinations of standard techniques. We note, however, that they all solve the more general problem of computing the whole intersection rather than just deciding whether it is nonempty. 

\begin{table}[t]
    \centering
    \renewcommand{\arraystretch}{1.7}
    \begin{NiceTabular}[hvlines,rules/color=[gray]{0.75}]{ c c | c c c c c c}
\CodeBefore
\rectanglecolor{red!15}{3-3}{3-8}
\rectanglecolor{green!15}{4-4}{4-4}
\rectanglecolor{red!15}{4-5}{4-5}
\rectanglecolor{green!15}{4-6}{4-8}
\rectanglecolor{green!15}{5-5}{5-8}
\rectanglecolor{green!15}{6-6}{6-8}
\rectanglecolor{green!15}{7-7}{7-8}
\rectanglecolor{green!15}{8-8}{8-8}
\Body
 \Block[]{2-2}{$(t,u)$-intersect.}&  & $ab^*a^*b$ & $[a|b][b|c]$ &  $ab^+a^+a$ &  $[aba | ba]$ & $[abba]^+$ & $[baab]^*$  \\
 & & $\conc *$ & $\conc |$ &  $\conc +$ & $|\conc $ & $+\conc$ & $*\conc$ \rule[-1ex]{0pt}{3ex}\\
    $ab^*a^*b$ &$\conc *$ & & & & & & \\
    $[a|b][b|c]$ & $\conc |$ & & & & & & \\
    $ab^+a^+a$ & $\conc +$ & & & & & & \\ 
    $[aba | ba]$ & $| \conc$ & & & & & & \\
    $[abba]^+$ & $+ \conc$ & & & & & & \\
   $[baab]^*$ & $* \conc$ & & & & & & 
\CodeAfter
\tikz{
\node (A1) at (3-|3)[below right=-3pt]{};
\node (B1) at (3-|9) [below left=-3pt]{};
\node (C1) at (4-|9) [above left=-3pt]{};
\node (D1) at (4-|3) [above right=-3pt]{};
\draw [myred,dashed,rounded corners=5pt] (A1.center) -- (B1.center) -- (C1.center) -- (D1.center) -- cycle;
\node at (3.5-|6) {\footnotesize {\bf Cor.~\ref{cor: conc_something-conc_star}} $\Omega((mn)^{1-\epsilon})$ \cite{focs16}};
}
\tikz{
\node (A1) at (4-|5)[below right=-3pt]{};
\node (B1) at (4-|6) [below left=-3pt]{};
\node (C1) at (5-|6) [above left=-3pt]{};
\node (D1) at (5-|5) [above right=-3pt]{};
\draw [myred,dashed,rounded corners=5pt] (A1.center) -- (B1.center) -- (C1.center) -- (D1.center) -- cycle;
\node[align=center,text width=35pt] at (4.5-|5.5) {{\bf\footnotesize  Thm.~\ref{thm:hardness}}  \\[-4pt] {\footnotesize $\!\!\Omega(\!(mn)^{1\!-\!\epsilon})$}};
}
\tikz{
\node (A1) at (4-|4)[below right=-3pt]{};
\node (B1) at (4-|5) [below left=-3pt]{};
\node (C1) at (5-|5) [above left=-3pt]{};
\node (D1) at (5-|4) [above right=-3pt]{};
\draw [mygreen,dashed,rounded corners=5pt] (A1.center) -- (B1.center) -- (C1.center) -- (D1.center) -- cycle;
\node[align=center,text width=35pt] at (4.5-|4.5) {{\bf\footnotesize  Thm.~\ref{thm:cdot-or}}  \\[-4pt] {\footnotesize $\mO(m+n)$}};
}
\tikz{
\node (A1) at (4-|6)[below right=-3pt]{};
\node (B1) at (4-|8) [below left=-3pt]{};
\node (C1) at (5-|8) [above left=-3pt]{};
\node (D1) at (5-|6) [above right=-3pt]{};
\draw [mygreen,dashed,rounded corners=5pt] (A1.center) -- (B1.center) -- (C1.center) -- (D1.center) -- cycle;
\node[align=center] at (4.5-|7) { \footnotesize {\bf  Thm.~\ref{thm:cdot-or}} $\mO(m+n)$};
}
\tikz{
\node (A1) at (5-|5)[below right=-3pt]{};
\node (B1) at (5-|8) [below left=-3pt]{};
\node (C1) at (6-|8) [above left=-3pt]{};
\node (D1) at (6-|5) [above right=-3pt]{};
\draw [mygreen,dashed,rounded corners=5pt] (A1.center) -- (B1.center) -- (C1.center) -- (D1.center) -- cycle;
\node[align=center] at (5.5-|6.5) {\footnotesize {\bf  Thm.~\ref{thm:cdot-plus}}  $\mO(m+n)$};
}
\tikz{
\node (A) at (6-|6)[below right=-3pt]{};
\node (B) at (6-|8) [below left=-3pt]{};
\node (C) at (8-|8) [above left=-3pt]{};
\node (D) at (8-|7) [above right=-3pt]{};
\node (E) at (7-|7) [above right=-2pt]{};
\node (F) at (7-|6) [above right=-2pt]{};
\draw [mygreen,dashed,rounded corners=5pt] (A.center) -- (B.center) -- (C.center) -- (D.center) -- (E.center) -- (F.center) -- cycle;
\node[align=center] at (6.5-|7) {\footnotesize {\bf  Thm.~\ref{thm:or-dot_plus-cot}}  $\mO(m+n)$};
}
\tikz{
\node (A1) at (4-|8)[below right=-3pt]{};
\node (B1) at (4-|9) [below left=-3pt]{};
\node (C1) at (9-|9) [above left=-3pt]{};
\node (D1) at (9-|8) [above right=-3pt]{};
\draw [mygreen,dashed,rounded corners=5pt] (A1.center) -- (B1.center) -- (C1.center) -- (D1.center) -- cycle;
\node[align=center,text width=35pt] at (6-|8.5) {{\bf\footnotesize  Thm.~\ref{thm:star-cdot}}  \\[-2pt] {\footnotesize $\mO(m+n)$}};
}
\end{NiceTabular}
    \caption{Classification of the time complexity of the $(t,u)$-intersection testing problem for types $t,u\in\{\conc*,\conc|,\conc+,|\conc,+\conc,*\conc\}$. The lower bounds assume SETH.}
    \label{tab:cdot_2deg_regexp} 
\end{table}

\begin{table}
 \renewcommand{\arraystretch}{1.1}
    \centering
    \begin{NiceTabular}[hvlines,rules/color=[gray]{0.75}]{ c c | c c c c c c}
\CodeBefore
    \rectanglecolor{green!15}{3-8}{14-8}
    \rectanglecolor{green!15}{3-7}{13-7}
    \rectanglecolor{green!15}{3-6}{12-6}
    \rectanglecolor{green!15}{3-5}{11-5}
    \rectanglecolor{green!15}{3-4}{10-4}
    \rectanglecolor{green!15}{3-3}{9-3}
\Body
 \Block[]{2-2}{$(t,u)$-intersect.}& & $[a^+|b^+|c]$ & $[a^*|b^*|c]$ &  $[a|b|c]^+$ &  $[a|b|c]^*$ & $[a^*]^+$ & $[a^+]^*$  \\
   &  & $|+$ & $|*$ &  $+|$ & $*|$ & $+* $ & $*+$ \rule[-1ex]{0pt}{3ex}\\ 
     $ab^*a^*b$ & $\conc *$ & & & & & & \\
      $[a|b][b|c]$& $\conc |$ & & & & & & \\
      $ab^+a^+a$&  $\conc +$ & & & & & & \\ 
     $[aba | ba]$ &  $| \conc$ & & & & & & \\
     $[abba]^+$ &  $+ \conc$ & & & & & & \\
     $[aab]^*$ &  $* \conc$ & & & & & & \\
     $[a^+|b^+|c]$ &  $|+$ & & & & & & \\
     $[a^*|b^*|c]$ &  $|*$ & & & & & & \\
     $[a|b|c]^+$ &  $+|$ & & & & & & \\
     $[a|b|c]^*$ &  $* |$ & & & & & & \\
     $[a^*]^+$ &  $+*$ & & & & & & \\
     $[a^+]^*$ &  $*+$ & & & & & & 
\CodeAfter
\tikz{
\node (A) at (3-|7) {};
\node (B) at (3-|9){}; 
\node (C) at (15-|9){};
\node (D) at (15-|8){};
\node (E) at (14-|8){};
\node (F) at (14-|7){};
\draw [mygreen,dashed,rounded corners=5pt] (A.south east) -- (B.south west) -- (C.north west) -- (D.north east) -- (E.north east) -- (F.north east) -- cycle;
\node[text width=40pt] at (7-|8) {{\bf Cor.~\ref{cor:star-plus}} \\$\mO(m+n)$};
\node[text width=40pt] at (7-|6) {{\bf Thm.~\ref{thm:star_plus-or}} \\$\mO(m+n)$};
\node[text width=40pt] at (7-|4) {{\bf Thm.~\ref{thm:or-star_plus}} \\$\mO(m+n)$};
}
\tikz{
\node (A) at (3-|5) {};
\node (B) at (3-|7){}; 
\node (C) at (13-|7){};
\node (D) at (13-|6){};
\node (E) at (12-|6){};
\node (F) at (12-|5){};
\draw [mygreen,dashed,rounded corners=5pt] (A.south east) -- (B.south west) -- (C.north west) -- (D.north east) -- (E.north east) -- (F.north east) -- cycle;
}
\tikz{
\node (A) at (3-|3) {};
\node (B) at (3-|5){}; 
\node (C) at (11-|5){};
\node (D) at (11-|4){};
\node (E) at (10-|4){};
\node (F) at (10-|3){};
\draw [mygreen,dashed,rounded corners=5pt] (A.south east) -- (B.south west) -- (C.north west) -- (D.north east) -- (E.north east) -- (F.north east) -- cycle;
}

\end{NiceTabular}
    \caption{Time complexity of $(t,u)$-intersection testing for all depth-$2$ types $t$ and $u\in\{|+,|*,+|,*|,+*,*+\}$.}
    \label{tab:test}
\end{table}

\noindent\textbf{Related work.}~
The more general problem of testing the intersection of $k\geq 2$ regular expressions can be solved in time proportional to the product of the sizes of the regexps by building the product automaton, and is known to be $\textbf{PSPACE}$-complete when the number of expressions and their sizes are unbounded~\cite{DBLP:conf/focs/Kozen77}. 
Recently, Su et al.~\cite{DBLP:conf/ictac/SuLPC23} and 
Chen et al.~\cite{chen2025incremental} proposed algorithms for this problem that avoid explicitly constructing the product automaton and often reduce the time and search space in practice. Bala~\cite{DBLP:conf/icalp/Bala02} showed that testing the intersection of an unbounded number of regexps remains \textbf{PSPACE}-complete when it is restricted to regexps without $+$ and with two nested stars, and it becomes \textbf{NP}-complete when there are no nested stars. Arrighi et al.~\cite{DBLP:conf/fsttcs/ArrighiF00JO021} investigated the complexity of intersection testing
for star-free language classes. Fernau et al.~\cite{DBLP:journals/algorithms/FernauK17} studied the complexity of intersection testing on finite automata parameterised by the alphabet size and the maximum number of states of the input automata. 
De Oliveira Oliveira and Wehar~\cite{10.1007/978-3-030-48516-0_6} proved that the intersection of two DFAs with $n$ states cannot be solved in $\mO(n^{2-\epsilon})$ for any $\epsilon>0$ assuming SETH. We remark that, since the automata constructed from regexps of types $\conc|$ or $\conc+$ are deterministic (modulo a simple linear-time transformation described in Theorem~\ref{thm:cdot-plus}), our conditional quadratic lower bound for $(\conc +,\conc |)$-intersection testing is a stronger result. 
Note that, in contrast, all the lower bounds in~\cite{focs16,DBLP:conf/focs/BringmannGL17} are for regexp types that give rise to NFAs.

The closely related problems of regular expression pattern matching and membership testing have been extensively investigated~\cite{DBLP:journals/jacm/Myers92,DBLP:conf/icalp/Bille06,DBLP:journals/tcs/BilleF08,DBLP:conf/icalp/BilleT09,DBLP:conf/soda/BilleT10,DBLP:conf/soda/BilleG24}, also in the streaming model~\cite{DBLP:conf/soda/DudekGGS22}. 
The classification of the time complexity of pattern matching and membership testing with homogeneous regular expressions of~\cite{focs16,DBLP:conf/focs/BringmannGL17} has later been refined by Abboud and Bringmann~\cite{DBLP:conf/icalp/AbboudB18} and Schepper~\cite{schepper:LIPIcs.ESA.2020.80}.
A great deal of work has also been carried out in devising efficient algorithms for pattern matching and intersection testing for specific kinds of homogeneous regexps, including depth-$2$ expressions of type $\conc |$ (also known in the literature as \emph{indeterminate, degenerate} or \emph{$1$-D} strings) or $| \conc$ (i.e. dictionaries of strings) and depth-$3$ expressions of type $\conc | \conc$ (known as \emph{Elastic-Degenerate} (ED) strings). 
Dictionary matching can be solved in linear time with the classic Aho-Corasick algorithm~\cite{DBLP:journals/cacm/AhoC75}; indeterminate string matching is well-studied~\cite{DBLP:conf/stoc/ColeH02,jda08,DBLP:journals/ipl/DaykinGGLLLMPW19,cpm20} and can be solved in time $\mO(n\log^2 m)$. To the best of our knowledge, indeterminate string intersection has not been considered explicitly in the literature, but a linear-time algorithm for intersecting the more general \emph{Generalised Degenerate} strings (concatenations of sets of strings of equal length) applies~\cite{fi20}. Several parameterised algorithms also exist to find occurrences of a string in an ED text~\cite{cpm17,cpm18,icalp19,sicomp22}. 
Note that the roles of the string and the regexp are reversed with respect to the usual regexp pattern matching; as observed in~\cite{DBLP:conf/wabi/Ascone0CEGGP24}, the quadratic conditional lower bound for membership testing of~\cite{focs16} implies a quadratic lower bound also for this version of the problem.
Intersection testing for ED strings has been considered in~\cite{cpm23} where, among other results, the authors showed SETH-hardness of the problem even when it is restricted to a binary alphabet.

\noindent\textbf{Paper organization.}~In \Cref{sec:preliminaries} we formally introduce the problem and the notation used throughout the paper and consider the intersection of a depth-$1$ with a depth-$2$ regexp. In \Cref{sec:main_reduction} we prove SETH-hardness for $(\conc +, \conc |)$-intersection testing. In \Cref{sec:upper_bounds} we present linear-time algorithms for all the other cases.

\section{Preliminaries}\label{sec:preliminaries}
A regular expression (regexp) over an alphabet $\Sigma$ and an operator set $O=\{\conc,|,+,*\}$ can be defined inductively as follows:
$c$ is a regular expression for any $c\in \Sigma$; given $\cA$ and $\cB$ two regular expressions, then $\cA\conc \cB$, $\cA | \cB$,  $\cA^+$ and $\cA^*$ are regular expressions. 
Each regular expression $\cA$ determines a language over $\Sigma$, denoted by $\cL(\cA)$.
In particular, given $c \in \Sigma$, $\cA$ and $\cB$ two regular expressions, we have: $\cL(c) = \{c\}$, $\cL(\cA\conc \cB) = \{w_1w_2 : w_1 \in \cL(\cA) ~\textnormal{and}~ w_2 \in \cL(\cB)\}$, $\cL(\cA| \cB) = \cL(\cA) \cup \cL(\cB)$, $\cL(\cA^+) = \{w_1\cdots w_k : k\ge 1 ~\textnormal{and}~ w_1,\dots, w_k\in \cL(\cA)\}$, $\cL(\cA^*) = \cL(\cA) \cup \{\epsilon\}$, where $\epsilon$ denotes the empty string. We will sometimes omit the symbol $\conc$ and simply write $\cA\cB$ to denote the concatenation $\cA\conc\cB$.

We denote by $\realph{\cA}\subseteq \Sigma$ the set of letters appearing in $\cA$. Given any regexp $\cA$ and integer $k>0$, $\cA^k$ denotes the concatenation of $k$ copies of $\cA$. For any $c\in\Sigma$, we call $c^k$ a \emph{run} of letter $c$. 
Throughout the paper, we assume that $\Sigma$ is an integer alphabet of polynomial size\footnote{Without this assumption, some of our upper bounds are increased by logarithmic factors. The same assumption was also implicitly made e.g. in~\cite[Lemma 4]{DBLP:conf/focs/BringmannGL17}
}, we denote by $\Sigma^*$ the set of all strings over $\Sigma$ (including $\epsilon$), and $\Sigma^+=\Sigma^*\setminus\{\epsilon\}$. Furthermore, we denote by $[k]$ the set of integers $\{1,2,\ldots,k\}$ and by $[k,\ell]$ any interval of integers.

Any regexp has a representation as a rooted tree with leaves labelled by letters from $\Sigma$ and inner nodes labelled by operators. 
Each inner node has one or multiple children, depending on the operator:
nodes labelled with $+$ or $*$ have exactly one child, while nodes labelled with $\conc$ or $|$ can have multiple children.
We define the \emph{size} of a regexp as the number of leaves in its tree representation\footnote{The size of a regexp is usually defined as the total number of nodes of the tree. We chose to count just the leaves to simplify some expressions across the paper: clearly, for trees of constant depth, the two quantities are asymptotically the same.
}.
A regular expression is \emph{homogeneous of type $t$ and depth $d$} if $t=t_1\cdots t_d\in \{\conc,|,+,*\}^d$, at any level $i$ of the tree representation all inner nodes are labelled with the operator $t_i$, and the depth of the tree is $d+1$. 
We say that the type of any inner node at level $i$ is $t_i$. 
Note that leaves can be at any level.
We assume that $t_i\ne t_{i+1}$; otherwise, we can merge the two levels into a single level. 
For example, the regexp $a^+b[c|a]$ is not homogeneous because both operators $+$ and $|$ label nodes at the first level of the tree representation;
examples of depth-2 homogeneous regexps of type $\conc +$ and $\conc |$ are in \Cref{fig:tree_example}.

The $(t,u)$\emph{-intersection testing} problem asks to decide if, given two homogeneous regular expressions of type $t$ and $u$, respectively, the intersection of the languages of the two expressions is nonempty.
Given two regular expressions $\cA,\cB$, we will use the shorthand $\cA\sqcap \cB$ to denote the language intersection $\cL(\cA)\cap\cL(\cB)$.

In this paper, we study the computational complexity of the $(t,u)$-intersection testing problem for all combinations of types $t,u$ of homogeneous regular expressions of depth 2.
For completeness, we briefly report the computational complexity of the intersection problem when one of the two homogeneous regexps is of depth 1. When the type of the depth-1 regexp is $\conc$ (thus it corresponds to a string), the intersection testing problem coincides with the membership testing problem, whose computational complexity has been fully determined for all kinds of homogeneous regexps by Backurs and Indyk~\cite{focs16} and Bringmann, Gr\o nlund and Larsen~\cite{DBLP:conf/focs/BringmannGL17}. When the depth-1 regexp is of type $|,+$ or $*$, the intersection problem with any other regular expression can be solved in linear time with folklore algorithms.

\begin{remark}\label{rk: depth-1}
Given a regular expression $\cA$ of depth 1 of type $|,+$ or $*$ and any other regular expression $\cB$, we can determine in linear time whether $\cA \sqcap \cB \ne \emptyset$.
\end{remark}

Thus, let us focus only on homogeneous regular expressions of depth 2. We first note that regular expressions of types $*+$ and $+*$ are equivalent to regexps of type $*$. Indeed, a regular expression of type $*+$ is of the form $(c^+)^* = (c^*)^+ = c^*$ for some $c\in \Sigma$. From \Cref{rk: depth-1}, we immediately obtain Corollary~\ref{cor:star-plus}.

\begin{cor}
\label{cor:star-plus}
    Problems $(*+,t)$-intersection testing and  $(+*,t)$-intersection testing can be solved in linear time for any type $t$ of homogeneous regexp.
\end{cor}

Backurs and Indyk~\cite{focs16} proved that, assuming SETH, the complexity of membership testing with a regexp of type $\conc *$ (a.k.a. $(\conc,\conc *)$-intersection) is $\Omega((mn)^{1-\alpha})$ for all $\alpha>0$. 
This result immediately implies the same conditional lower bound for the time complexity of the intersection testing problem for the cases stated in Corollary~\ref{cor: conc_something-conc_star}.

\begin{cor}\label{cor: conc_something-conc_star}
  Problem $(\conc *,t)$-intersection testing for all types $t\in \{\conc *,\conc |, \conc +,|\conc,\\ +\conc, *\conc\}$ cannot be solved in time $\mO((mn)^{1-\alpha})$ for any $\alpha>0$ unless SETH fails.
\end{cor}

\section{SETH-hardness of ($\conc+,\conc|$)-intersection testing}\label{sec:main_reduction}
 Let $\cA$ be of type $\conc +$ and size $m$ and $\cB$ be of type $\conc |$ and size $n$. We define a \emph{position} in $\cA$ or $\cB$ as a node on the first level of their tree representation. Thus, for any position $i$, $\cA[i]$ is either a letter $a$ or $a^+$; $\cB[i]$ is either a single letter or a set of letters.
A \emph{fragment} $\cA[i\dd j]$ is the regular expression given by the concatenation of the positions $\cA[i]\cA[i+1]\cdots\cA[j]$ (a fragment of $\cB$ is defined analogously). 
An \emph{alignment} of a string $S\in\cL(\cA)$ with $\cA$ is a monotonic surjective mapping $\phi_{\cA}$ 
of the positions of $S$ to the positions of $\cA$ (i.e. $i<j\Rightarrow \phi_{\cA}(i)\leq \phi_{\cA}(j)$~) such that $S[i\dd j]\in \cL(\cA[\phi_{\cA}(i)\dd \phi_{\cA}(j)])$ for all $i\leq j$. Let $k\leq n$ be the number of positions of $\cB$: since all the strings $S\in\cL(\cB)$ have length $k$, the alignment of $S$ with $\cB$ is defined by the identity function, that is, $\phi_{\cB}(i)=i$ for all $i\in[k]$. 
An alignment of $\cB$ with $\cA$ is defined as a monotonic surjective mapping $\psi$ of the positions of $\cB$ to the positions of $\cA$ such that $\cB[i\dd j]\sqcap \cA[\psi(i)\dd \psi(j)]\neq\emptyset$ for all $i\leq j\leq k$.
We say that a fragment $\cB[i\dd j]$ and a fragment $\cA[i'\dd j']$ are aligned if $i'=\psi(i)$ and $j'=\psi(j)$. 

\begin{figure}[ht]
    \centering
    \includegraphics[width=\textwidth]{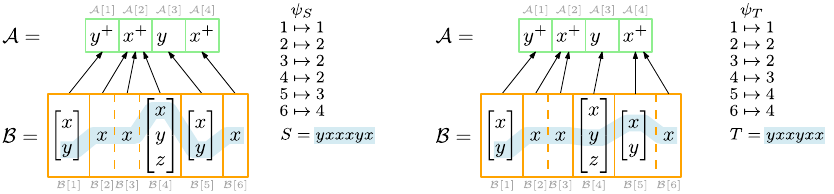}
    \caption{Two alignments of $\cA=y^+x^+yx^+$ and $\cB=[x|y]xx[x|y|z][x|y]x$, respectively corresponding to strings $S=yxxxyx$ (left) and $T=yxxyxx$ (right), both highlighted in $\cB$. Dashed lines separate positions of  $\cB$ that are mapped to the same position of $\cA$.}\label{fig:alignment_example}
\end{figure}

\begin{observation}\label{obs:alignment_gives_intersection}
    Any alignment $\psi$ of $\cB$ with $\cA$ corresponds to a string $S\in \cA \sqcap \cB$. This is because $\cB[1\dd k]\sqcap \cA[\psi(1)\dd \psi(k)]\neq\emptyset$ and since $\psi$ is surjective, this is equal to $\cB \sqcap \cA$; and in particular, we have that $S[i]=c_i$ if $\cA[\psi(i)]=c_i$ or $\cA[\psi(i)]=c_i^+$.
\end{observation}
Conversely, note that any string $S\in \cA \sqcap \cB$ must have length $k$ and must align with both $\cA$ and $\cB$: in this case, the alignment $\phi_\cA$ of $S$ with $\cA$ induces an alignment $\psi_S$ of $\cB$ with $\cA$ such that $\psi_S(i)=\phi_\cA(i)$ for all $i\in[k]$. See \Cref{fig:alignment_example} for an example.

We next show a quadratic lower bound for $(\conc+,\conc |)$-intersection testing with a reduction from Orthogonal Vectors (\ov). 

\begin{definition}[Orthogonal Vectors]
\label{def:ov}
    Given two sets $A, B \subseteq \{ 0,1 \}^d$ such that $|A| = M$, $ |B| = N$ determine whether there exist $\alpha \in A$ and $\beta \in B$ such that $\alpha$ and $\beta$ are orthogonal, namely, $\alpha \cdot \beta = \sum_{i=1}^{d} \alpha[i] \cdot \beta[i] = 0$.
\end{definition}
Our reduction relies on the following restricted case of the Unbalanced Orthogonal Vectors Hypothesis (\uovh), which is known to be implied by SETH~\cite{10.1109/FOCS.2015.15}.
\begin{conjecture*}[Unbalanced Orthogonal Vectors Hypothesis]
\label{def:ovh}
    For no $\epsilon > 0$ there is an algorithm for \ov, restricted to $M= \Theta(N)$ and $d  \le N^{o(1)}$, that runs in time $\mO(N^{2-\epsilon})$.
\end{conjecture*}

\noindent\textbf{Assumptions on the \ov instance.}~
We make the following assumptions on the \ov instance $A=\{\alpha_1,\ldots,\alpha_M\}, B=\{\beta_1,\ldots,\beta_N\}$ with $M=\Theta(N)$, which can all be ensured via trivial linear-time modifications to any general instance.

\begin{enumerate}
    \item\label{ov: M_odd_N_even} $M$ is an odd integer, $N$ is an even integer, and $M< N$; furthermore, 
    $N \equiv 0 (\textnormal{mod} ~4)$
    (can be obtained by adding multiple copies of some vectors of $A$ and/or $B$).
    \item\label{ov: odd} The length $d$ of the vectors is odd (can be obtained by adding a $0$ at the end of all vectors in case they have even length).
    \item\label{ov: A_zeros_B_ones} All vectors of set $A$ begin and end with a $1$; all vectors of $B$ begin and end with a $0$ (can be obtained by appropriately padding all the vectors at the beginning and the end).
    \item\label{ov: no_trivial_vectors} $A$ and $B$ do not contain the vectors $1^d$, $0^d$, $10^{d-2}1$, and $01^{d-2}0$ (can be checked with a linear-time scan).
    \item\label{ov: a1_aM_not_orthogonal} $\alpha_1\notperp \beta_j$ and $\alpha_M\notperp \beta_j$ for all $j\in [N]$.
    \item\label{ov: ai_bj_i_j_same_parity} If there are $i,j$ such that $\alpha_i \cdot \beta_j = 0$ then there are $i',j'$ such that $\alpha_{i'} \cdot \beta_{j'} = 0$ and $i'\equiv j' (\textnormal{mod}~2)$ (Assumption $3$ in~\cite{focs16} justifies why this can be assumed w.l.o.g.).
\end{enumerate}

\noindent\textbf{Gadgets.}~Our reduction uses the constant-size alphabet $\Sigma = \{x,y,\$\}$.
We begin by showing how to encode the vectors of $A$ with regexp gadgets of type $\conc +$. 
Coordinate gadgets encode the entries of a vector with the following rationale: even positions correspond to gadgets that only use the letter $x$, while the gadgets for the odd positions only use $y$. 
The size of a gadget is determined by the value of the corresponding entry in the vector, namely, size-3 gadgets of the form $ccc^+$ encode entries equal to 1, and size-1 gadgets of the form $c^+$ encode $0$ entries (where $c\in \{x,y\}$ according to the parity of the position of the vector). Formally, we thus define coordinate gadgets $C_A(v,k)$ for $v\in\{0,1\}$ and $k\in[d]$ as follows:

\[C_A(v,k)\coloneq
\begin{cases}
    yyy^+ & \textnormal{for} ~v=1 ~\textnormal{and}~ k\equiv 1 (\textnormal{mod} ~2)\\
    xxx^+ & \textnormal{for} ~v=1 ~\textnormal{and}~ k\equiv 0 (\textnormal{mod} ~2)\\
    y^+ & \textnormal{for} ~v=0 ~\textnormal{and}~ k\equiv 1 (\textnormal{mod} ~2)\\
    x^+ & \textnormal{for} ~v=0 ~\textnormal{and}~ k\equiv 0 (\textnormal{mod} ~2).
\end{cases}\]

Vector gadgets $a_i^\perp$ are then defined as the concatenation of the coordinate gadgets for $\alpha_i$: \[a_i^\perp \coloneq C_A(\alpha_i[1],1)C_A(\alpha_i[2],2)\cdots C_A(\alpha_i[d],d) .\]

Due to Assumptions~\ref{ov: odd} and~\ref{ov: A_zeros_B_ones}, all such vector gadgets begin with $yyy^+$ and end with $yyy^+$.  
To encode the vectors of $B$, we define coordinate gadgets which still associate letter $y$ with odd positions and letter $x$ with even positions, but this time size-$1$ gadgets encode the occurrences of $1$, and size-$3$ gadgets the occurrences of $0$; since $\cB$ is of type $\conc |$, the gadgets do not contain the operator $+$ and the vectors of $B$ are encoded with solid strings.  
Formally, we define the coordinate gadgets $C_B(v,k)$ for $v\in\{0,1\}$ and $k\in[d]$ as follows:
\[C_B(v,k)\coloneq
\begin{cases}
    y & \textnormal{for} ~v=1 ~\textnormal{and}~ k\equiv 1 (\textnormal{mod} ~2)\\
    x & \textnormal{for} ~v=1 ~\textnormal{and}~ k\equiv 0 (\textnormal{mod} ~2)\\
    yyy & \textnormal{for} ~v=0 ~\textnormal{and}~ k\equiv 1 (\textnormal{mod} ~2)\\
    xxx & \textnormal{for} ~v=0 ~\textnormal{and}~ k\equiv 0 (\textnormal{mod} ~2).
\end{cases}\]

Vector gadgets $b_j$ are then defined depending on the parity of $j$ (given a fixed but arbitrary ordering of the vectors of $B$):
\[b_j\coloneq 
\begin{cases}
    C_B(\beta_j[1],1)C_B(\beta_j[2],2)\cdots C_B(\beta_j[d],d) &\textnormal{if} ~j \equiv 1 (\textnormal{mod} ~2)\\
    yyyC_B(\beta_j[1],1)C_B(\beta_j[2],2)\cdots C_B(\beta_j[d],d)yyy &\textnormal{if} ~j \equiv 0 (\textnormal{mod} ~2).
\end{cases}\]

Due to Assumptions~\ref{ov: odd} and~\ref{ov: A_zeros_B_ones}, all gadgets $b_j$ with an odd $j$ start and end with $y^3$, and those with an even $j$ start and end with $y^6$. 
We also define another two gadgets of type $\conc +$
\begin{equation}\label{eq: a_zero_a_even}
a_0^\perp\coloneq (y^+x^+)^{\lfloor\frac{d}{2}\rfloor}y^+,
~~~~~~~~
a_{even}^\perp\coloneq y^6x^+(y^+x^+)^{\lfloor\frac{d}{2}\rfloor-1}y^6
\end{equation}
and three additional gadgets of type $\conc |$ 
\begin{equation}\label{eq: b_zero_dollar}
b_0\coloneq \left(y^3\left[x|
y\right]^3\right)^{\lfloor\frac{d}{2}\rfloor}y^3; ~~~~~ b_{even}\coloneq y^6x\left(yx\right)^{\lfloor\frac{d}{2}\rfloor-1}y^6; ~~~~~ b_{odd}\coloneq y^3x\left(yx\right)^{\lfloor\frac{d}{2}\rfloor-1}y^3.
\end{equation}
Finally, we define $b_0^{(\$)} \coloneq b_0 \left[y|\$\right]$, $ b_{even}^{(\$)} \coloneq b_{even}
\left[y|\$\right]$, and $b_{odd}^{(\$)} \coloneq b_{odd} \left[y|\$\right]$. 

\begin{example}\label{ex:gadgets}
    Let $d=5$, consider the vectors $\alpha_1=10011$, $\alpha_2=11001$, $\beta_1=00010$, $\beta_2=01010$, and denote by $x^{3+}$ and $y^{3+}$ the expressions $xxx^+$ and $yyy^+$, respectively. 
    The associated gadgets are $a_1^\perp=y^{3+}x^{+}y^+x^{3+}y^{3+}$; $a_2^\perp=y^{3+}x^{3+}y^{+}x^{+}y^{3+}$; $b_1=y^3x^3y^3xy^3$; $b_2=y^6xy^3xy^6$. 
    Note that $a_2^\perp\sqcap b_1 \neq \emptyset$, but $a_1^\perp\sqcap b_1= \emptyset$: indeed, $\beta_1$ is orthogonal to $\alpha_2$, but not to $\alpha_1$.
    Furthermore, $a_0^\perp=y^+x^+y^+x^+y^+$; $a_{even}^\perp=y^6x^+y^+x^+y^6$; $b_0=y^3\left[x|y\right]^3 y^3 \left[x|y\right]^3 y^3$; $b_0^{(\$)}=y^3\left[x|y\right]^3 y^3 \left[x|y\right]^3 y^3 \left[y|\$\right]$; $b_{even}^{(\$)}=y^6xyxy^6\left[y|\$\right]$; and $b_{odd}^{(\$)}=y^3xyxy^3\left[y|\$\right]$.
\end{example}

\Cref{lem: orthogonal_gadgets} is straightforward from the definitions of the vector gadgets $a_i^\perp$ and $b_j$.

\begin{lemma}\label{lem: orthogonal_gadgets}
    For any $i\in[M]$ and $j\in [N]$, $a_i^\perp \sqcap b_j\ne \emptyset$ (thus $a_i^\perp$ aligns with $b_j$) if and only if $\alpha_i\perp \beta_j$.
\end{lemma}

Since gadgets $b_j$ are strings, we have $a_i^\perp \sqcap b_j\ne \emptyset$ if and only if $b_j \in \cL(a_i^\perp)$, 
thus $\cL(a_i^\perp)$ contains the string $b_j$ for each vector $\beta_j$ orthogonal to $\alpha_i$.
We remark that, morally, $a_0^\perp$ and $b_0$ encode the vector $0^d$, while $b_{odd}$ and $b_{even}$ encode the vector $0 1^{d-2}0$. 
This is consistent with the fact that (i) $b_0 \in \cL(a_i^\perp)$ and $b_j \in \cL(a_0^\perp)$ for all $i\in [M],j\in [N]$ (as $0^d$ is orthogonal to any vector), and (ii) $a_i^\perp$ does not align with $b_{odd}$ or $b_{even}$ 
(as by Assumption~\ref{ov: no_trivial_vectors}, $A$ does not contain the vector $10^{d-2}1$, and thus $0 1^{d-2}0 \notperp \alpha_i$ for all $i\in[M]$). Furthermore, since $\cL(b_0)$ contains a run of the letter $y$, we can align $a_0^\perp$ with $b_0^k$ for any $k\ge1$.
We summarise these properties, which are straightforward to verify from the gadget definitions, in \Cref{rk: gadgets_matches}.

\begin{remark}\label{rk: gadgets_matches}
    The following relations among the gadgets hold:
    \begin{enumerate}
        \item \label{a_zero_b_i}  $a_0^\perp \sqcap (b_0)^k b_j\neq\emptyset$ for all $j\in[N]$ for all $k\ge 0$.
        \item\label{b_zero_a_i} $a_i^\perp \sqcap (b_0)^k\neq\emptyset$ for all $i\in[0,M]$ for all $k\ge 1$.   
         \item\label{b_0-b_j_a_i_ortho} $a_i^\perp \sqcap (b_0)^kb_j\neq\emptyset$ if and only if $\alpha_i\perp\beta_j$  for all $k\ge 0$.
        \item\label{b_even_odd_not_a_i} 
        $a_0^\perp \sqcap (b_0)^k b_{even} (b_0)^h\neq\emptyset$ and $a_0^\perp \sqcap (b_0)^k b_{odd} (b_0)^h\neq\emptyset$, but
        $a_i^\perp \sqcap (b_0)^k b_{even} (b_0)^h=\emptyset$ and $a_i^\perp \sqcap (b_0)^k b_{odd} (b_0)^h=\emptyset$ for all $i\in[M]$, for all $h,k\ge0$.
        \item\label{a_even_b_even_b_odd} $a_{even}^\perp \sqcap b_{even}\neq\emptyset$ but  $a_{even}^\perp \sqcap b_0=\emptyset$ and $a_{even}^\perp \sqcap b_{odd}=\emptyset$.

        \item\label{a_even_b_p_even} $a_{even}^\perp \sqcap b_p \ne \emptyset$ if and only if $p$ is even.

        \item \label{a_0_cannot_eat_b_even_b_odd} Since there are exactly $d-1$ alternation of $y$ and $x$ in $a_i^\perp$ for $i\in [0,M]$, $a_i^\perp$ can align with at most one string $b_j$ (including $b_{even}$ or $b_{odd}$ if $i=0$); for instance, $a_0^\perp \sqcap b_{even} b_0 b_{odd}=\emptyset$.
    \end{enumerate}
\end{remark}
For all cases of \Cref{rk: gadgets_matches}, given a regexp $\cA$ of type $\conc +$ and $\cB$ of type $\conc |$, the following are equivalent: $\cA \sqcap \cB \ne \emptyset$, $\cA \$\sqcap \cB\$ \ne \emptyset$, $\$ \cA \sqcap \$ \cB \ne \emptyset$, $\$ \cA\$\sqcap \$ \cB \$\ne \emptyset$.
This holds even if the symbol $\$$ is replaced by $[y|\$]$ in the regexp of type $\conc |$, or $b_0$ is replaced by $b_0^{(\$)}$.
\medskip

\noindent\textbf{Full reduction.}~
We concatenate the gadgets encoding \ov sets $A$ and $B$ to construct two regexps $\cA$, $\cB$ of types $\conc +$ and $\conc |$, respectively. 
Each regexp is partitioned into three fragments: $\cA=\cA_{pre}\cA^\perp\cA_{suf}$ and $\cB=\cB_{pre}\cB^\perp\cB_{suf}$.
The $pre$ and $suf$ fragments are designed to force the alignment of a fragment of  $\cA^\perp$ with a fragment of $\cB^\perp$ whenever we have a global alignment of $\cA$ and $\cB$, which in turn will imply that a pair of orthogonal vectors exists.

$\cA^\perp$ and $\cB^\perp$ are the central fragments of $\cA$ and $\cB$, respectively. They contain the gadgets $a_i^\perp$ and $b_j$, which do not appear anywhere else in $\cA$ and $\cB$, and are defined as follows.
\begin{align}
    \label{eq: calA_perp}
   \cA^\perp &\coloneq a_1^\perp\$ a_0^\perp\$ a_2^\perp\$ \cdots \$a_{M-1}^\perp\$a_0^\perp\$ a_M^\perp
   \\[2pt] \label{eq: calB_perp}
   \cB^\perp &\coloneq b_0^{(\$)} b_1 \$ b_0^{(\$)} b_2 \$ \cdots \$b_0^{(\$)} b_{N-1} \$b_0^{(\$)} b_N \$ .
\end{align}
\vspace{-12pt}
\begin{remark}\label{rk: number_of_dollars_cAperp_cBperp}
    All strings in the language of $\cA^\perp$ contain $2M-2$ occurrences of $\$$; all strings in the language of $\cB^\perp$ contain at least $N$ and at most $2N$ occurrences of $\$$.
\end{remark}

\begin{figure}[!ht]
    \centering
    \includegraphics[width=\textwidth]{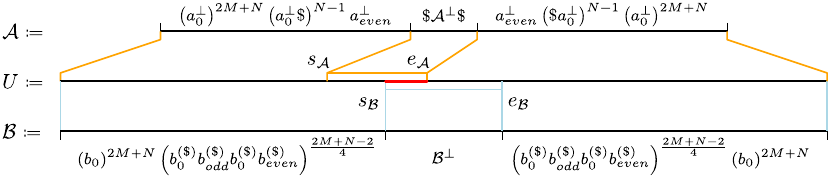}
    \caption{Visual representation of \Cref{lem: calA_perp_intersects_calB_perp}. The blue lines are vertical to emphasise that the alignment of $U$ with $\cB$ is the identity function.}  \label{fig:intersection_intervals}
\end{figure}

The two regular expressions $\cA$ and $\cB$ are then defined as

\begin{align}
    \label{eq: calA}
    \cA &\coloneq 
    \underbracket{\left(a_0^\perp\right)^{2M+N}\left(a_0^\perp\$\right)^{N-1} a_{even}^\perp\$}_{\cA_{pre}}  \cA^\perp \underbracket{\$a_{even}^\perp
    \left(\$a_0^\perp\right)^{N-1}\left(a_0^\perp\right)^{2M+N}}_{\cA_{suf}}
    \\[5 pt]
    \label{eq: calB}
    \cB &\coloneq 
    \underbracket{\left(b_0\right)^{2M+N}\left(b_0^{(\$)}b_{odd}^{(\$)}b_0^{(\$)}b_{even}^{(\$)}\right)^{\frac{2M+N-2}{4}}}_{\cB_{pre}} 
    \cB^\perp 
    \underbracket{\left(b_0^{(\$)}b_{odd}^{(\$)}b_0^{(\$)}b_{even}^{(\$)} \right)^{\frac{2M+N-2}{4}}\left(b_0\right)^{2M+N}}_{\cB_{suf}}.
\end{align}
\vspace{-12pt}
\begin{remark}\label{rk: A_B_fixed_dollars}
    All strings in the language of $\cA$ have exactly $2M+2N-2$ occurrences of \$, and
    all strings in the language of $\cB$ have at least $N$ and at most $4N+4M-4$ 
    occurrences of $\$$.
\end{remark}
Recall that by Assumption~\ref{ov: M_odd_N_even}, $N\equiv0 (\textnormal{mod}~4)$ and $2M \equiv 2 (\textnormal{mod}~4)$ thus $2M+N-2 \equiv 0 (\textnormal{mod}~4)$. 
Suppose that the intersection of the languages of $\cA$ and $\cB$ is not empty and consider a string $U\in \cA \sqcap \cB$. By Remark~\ref{rk: A_B_fixed_dollars}, $U$ has exactly $2M+2N-2$ occurrences of \$.
Lemma~\ref{lem: calA_perp_intersects_calB_perp} shows that the fragment $U[s_{\cA}\dd e_{\cA}]$ aligned with $\$\cA^\perp\$$ and the fragment $U[s_{\cB}\dd e_{\cB}]$ aligned with $\cB^\perp$ must overlap, and thus a fragment of $\cA^\perp$ aligns with a fragment of $\cB^\perp$: see also \Cref{fig:intersection_intervals}.

\begin{lemma}\label{lem: calA_perp_intersects_calB_perp}
    For any $U\in \cA \sqcap \cB$, let $U[s_{\cA}\dd e_{\cA}]$ and $U[s_{\cB}\dd e_{\cB}]$ be the fragments aligned with $\$\cA^\perp\$$ and $\cB^\perp$, respectively.
    Then  $[s_{\cA},e_{\cA}] \cap [s_{\cB
},e_{\cB}] \neq \emptyset$.  
\end{lemma}

\begin{proof}
By the definition of $\cA$, $s_\cA$ and $e_\cA$ are the positions of the $N$-th and the $(N+2M-1)$-th occurrence of \$  in $U$, respectively.
Thus the only way $[s_{\cA},e_{\cA}] \cap [s_{\cB
},e_{\cB}]$ could be empty is if the first $N+2M-1$ occurrences of \$ in $U$ are all all aligned within $\cB_{pre}$ or symmetrically, if the last $N+2M-1$ occurrences are all aligned within $\cB_{suf}$.
A simple counting argument shows that this is impossible. 
In fact, by the definition of $\cB$, at most $N+2M-2$ occurrences of \$ can be aligned within $\cB_{pre}$; and symmetrically, within $\cB_{suf}$. 
\end{proof}

\begin{lemma}
    If $\cA \sqcap\cB \ne \emptyset$ then there exists a pair of orthogonal vectors.
\end{lemma}
    \begin{figure}[t]
        \centering
        \includegraphics{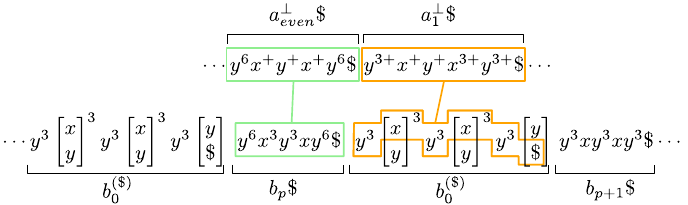}
        \caption{Example of the alignment of $a_{even}^\perp \$a_{1}^\perp\$$ with $b_{p}\$ b_{0}^{(\$)}$, for an even $p$. Here $d=5$, $\alpha_1= 10011$, $\beta_p=00010$, and $\beta_{p+1}=01010$. Note that, coherently with Assumption~\ref{ov: a1_aM_not_orthogonal}, $a_1^\perp$ does not align with $b_p$ nor with $b_{p+1}$.}
        \label{fig:a_even_sqcap_b_p_second_case}
    \end{figure}

\begin{proof}
    
Since $\cA \sqcap\cB\ne \emptyset$, $\cA$ can be aligned with $\cB$. The general idea is to prove that for each possible alignment of $\cA$ and $\cB$ there exist $i\in[M],j\in[N]$ such that $a_i^\perp$ is aligned with $b_0^{(\$)}b_j$ or $b_j$, which by \Cref{rk: gadgets_matches}.\ref{b_0-b_j_a_i_ortho} and \Cref{lem: orthogonal_gadgets} implies that $\alpha_i \perp \beta_j$. 
Let the intervals $[s_{\cA},e_{\cA}]$ and $[s_{\cB},e_{\cB}]$ be defined as in Lemma~\ref{lem: calA_perp_intersects_calB_perp}.
Since the intersection of these two intervals is not empty, we study the following three cases.
 
\begin{description}
    \item[Case $s_{\cB} \le s_\cA < e_\cA \le e_{\cB}$.]
    Let us focus on how the extremities of $\$\cA^\perp\$$, that is, $\$ a_1^\perp $ and $a_M^\perp\$$, can be aligned inside $\cB^\perp$. 
    In principle, the first \$ could be either aligned with an explicit occurrence of \$ in $\cB^\perp$ or with an ``optional'' occurrence in a set $[y|\$]$.
    The second case would imply that $\$ a_1^\perp \$$ aligns with $[y|\$] b_j\$$ for some $j\in [N]$, which is a contradiction due to Assumption~\ref{ov: a1_aM_not_orthogonal}.
    Therefore, it must be aligned with an explicit occurrence of \$ in $\cB^\perp$, which implies, in particular, that the whole $a_{even}^\perp\$a_1^\perp$ must be aligned within $\cB^\perp$.
    Again by Assumption~\ref{ov: a1_aM_not_orthogonal}, $\$a_1^\perp\$$ cannot align with $\$b_0^{(\$)}b_j\$$ for any $j\in[N]$;
     furthermore, by \Cref{rk: gadgets_matches}.\ref{a_even_b_p_even}, $a_{even}^\perp\$$ aligns with $ b_p \$$ if and only if $p$ is even. 
     The whole reasoning implies that  $a_{even}^\perp \$ a_1^\perp\$ $ must be aligned with $b_p\$b_0^{(\$)}$ for some even $p$ (see \Cref{fig:a_even_sqcap_b_p_second_case}).
    A similar reasoning proves that $a_M^\perp \$a_{even}^\perp\$$ must be aligned with $b_0^{(\$)}b_q\$$ for some even $q$.

    Let us study what happens in the middle part of $\cA^\perp$.
    Fixed any $i\in [2,M-1]$,  $\$a_i^\perp \$$ can only align with the following types of fragments of $\cB^\perp$: $\$ b_0^{(\$)}$ or $\$b_0^{(\$)}b_j\$$ or $[y|\$]b_j \$$. By \Cref{lem: orthogonal_gadgets} and \Cref{rk: gadgets_matches}.\ref{b_0-b_j_a_i_ortho}, the last two cases can occur if and only if $\alpha_i \perp \beta_j$.
    In contrast, $\$ a_0^\perp\$$ can align with  $\$b_0^{(\$)}$ or $\$ b_0^{(\$)}b_j\$$ or $[y|\$]b_j \$$ without any implications on vector orthogonality.
    Now since $a_{even}^\perp \$ a_1^\perp\$$ is aligned with $b_p\$b_0^{(\$)}$ for some even $p$,  $a_{even}^\perp \$ a_1^\perp\$ a_0^\perp\$$ must be aligned with $b_p\$b_0^{(\$)}b_{p+1}\$$.
    Suppose for a contradiction that $\alpha_i \not \perp \beta_{p+i}$ for all $i\in [M]$.
    Then $\alpha_2 \not \perp \beta_{p+2}$ implies that $a_{even}^\perp \$ a_1^\perp\$ a_0^\perp\$ a_2^\perp\$a_0^\perp\$$ must be aligned with $b_p\$b_0^{(\$)}b_{p+1}\$b_0^{(\$)}b_{p+2}\$$, and in particular, $b_0^{(\$)}$ is aligned with $a_2^\perp\$$ and $b_{p+2}\$$ is aligned with $a_0^\perp\$$.
    Iterating this process, we get that $a_{even}^\perp \$ a_1^\perp\$ a_0^\perp\$ a_2^\perp\$a_0^\perp\$\cdots \$ a_{M-1}^\perp\$a_0^\perp\$a_{M}^\perp \$$ is aligned with $b_p\$b_0^{(\$)}b_{p+1}\$b_0^{(\$)}b_{p+2}\$\cdots \$b_0^{(\$)}b_{p+M-1}\$b_0^{(\$)}$. 
    This leads to a contradiction because it would force $a_{even}^\perp\$$ to align with $b_{p+M}\$$, which is impossible since $p+M$ is odd by Assumption~\ref{ov: M_odd_N_even}. 
    We conclude that a pair of orthogonal vectors exists.
    \begin{center}
        \includegraphics{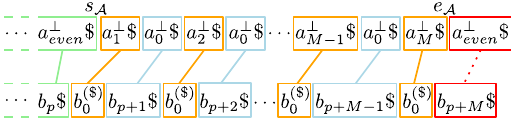}
    \end{center}
     \item[Case $s_\cA < s_{\cB}$.]
     Since $\cB[s_{\cB}] = y$ and $\cA[e_{\cA}] = \$$, we have $s_{\cB} \ne e_\cA$.
    Since $s_\cA < s_{\cB}$, the fragment $ a_{even}^\perp \$ a_1^\perp \$$ must be aligned with some fragment of $\cB$ starting to the left of $\cB^\perp$, i.e. in $\cB_{pre}$. 
    In particular, $a_{even}^\perp \$ a_1^\perp \$ $ can only align with some occurrence of $b_{even}^{(\$)}b_0^{(\$)}$ (in the case where $s_\cA=s_{\cB}-1$, it is aligned with the last occurrence of $b_{even}^{(\$)}$ in $\cB_{pre}$ and the first occurrence of $b_0^{(\$)}$ within $\cB^\perp$). Indeed, by Remark~\ref{rk: gadgets_matches}.\ref{a_even_b_even_b_odd}, $a^{\perp}_{even} \$ $ can only align with $b_{even}\$ $, and by Remark~\ref{rk: gadgets_matches}.\ref{b_even_odd_not_a_i}, $a_1^\perp$ does not align with $b_0^{(\$)}b_{odd}$: see \Cref{fig:a_even_sqcap_b_even}. 
    \begin{figure}[t]
        \centering
        \includegraphics{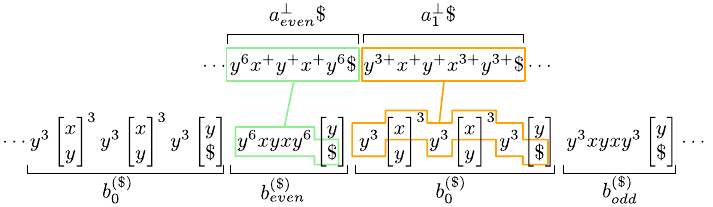}
        \caption{Example of the alignment $a_{even}^\perp \$a_{1}^\perp\$$ and $b_{even}^{(\$)}b_{0}^{(\$)}$ with $d=5$ and $\alpha_1= 10011$.}
        \label{fig:a_even_sqcap_b_even}
    \end{figure}
    In principle, $a_{even}^\perp \$ a_1^\perp \$a_0^\perp\$$ could be aligned with  $b_{even}^{(\$)}b_0^{(\$)} b_{odd}^{(\$)}b_0^{(\$)}$, and in particular, the last two positions of $b_{odd}^{(\$)}$ and the whole $b_0^{(\$)}$ (except its last position) could be aligned with the last position of $a_0^\perp$ (which corresponds to $y^+$: see Equations~\ref{eq: a_zero_a_even}, \ref{eq: b_zero_dollar}). 
    However, this would imply that $a_2^\perp\$$ would have to align with $b_{even}^{(\$)}$, which is not possible by Remark~\ref{rk: gadgets_matches}.\ref{b_even_odd_not_a_i}.
    By Remark~\ref{rk: gadgets_matches}.\ref{a_0_cannot_eat_b_even_b_odd}, $a_{even}^\perp \$ a_1^\perp \$a_0^\perp\$$ cannot be aligned with $b_{even}^{(\$)}b_0^{(\$)} b_{odd}^{(\$)}b_0^{(\$)}b_{even}^{(\$)}$ or any longer fragment;   
    therefore $a_{even}^\perp \$ a_1^\perp \$a_0^\perp\$$ must be aligned with $b_{even}^{(\$)}b_0^{(\$)} b_{odd}^{(\$)}$ so that $a_2^\perp \$$ can be aligned with $b_0^{(\$)}$.
    Proceeding with the alignment and iterating this reasoning, we get that for any odd $i$ such that $a_i^\perp$ is aligned within $\cB_{pre}$, $a_i^\perp \$a_0^\perp \$ a_{i+1}^\perp \$a_0^\perp \$$ must be aligned with $b_0^{(\$)} b_{odd}^{(\$)} b_0^{(\$)} b_{even}^{(\$)}$.
    
    Therefore, for any even $p$,
    $a_{even}^\perp \$ a_1^\perp \$a_0^\perp\$\cdots \$a_p^\perp\$a_0^\perp \$$ must be aligned with either (i) $b_{even}^{(\$)}(b_0^{(\$)} b_{odd}^{(\$)} b_0^{(\$)} b_{even}^{(\$)})^{\frac{p}{2}}$ or (ii) $b_{even}^{(\$)}(b_0^{(\$)} b_{odd}^{(\$)} b_0^{(\$)} b_{even}^{(\$)})^{\frac{p}{2}}b_0^{(\$)}$, implying that the smallest $i$ such that $a_i^\perp$ is aligned with a fragment of $\cB^\perp$ is always odd.
    Case (ii) occurs if and only if the last $a_0^\perp\$$ in $a_{even}^\perp \$ a_1^\perp \$a_0^\perp\$\cdots \$a_p^\perp\$a_0^\perp \$$ aligns with $b_{even}^{(\$)}b_0^{(\$)}$. This forces $a_{p+1}^\perp\$$ to align with $b_1\$$, which implies that $\alpha_{p+1} \perp \beta_1$ by \Cref{lem: orthogonal_gadgets}. So, suppose we are in Case (i).
    We now study the alignment of $ a_{p+1}^\perp \$a_0^\perp\$\cdots \$a_0^\perp\$a_M^\perp \$ $ (that is the remaining part of $\cA^\perp$). 
    Suppose for the sake of contradiction that $\alpha_{p+i} \not \perp \beta_{i}$ for all $i\in [1,M-p]$.
    Since $\alpha_{p+1} \not \perp \beta_{1}$, then $ a_{p+1}^\perp \$$ can only be aligned with $b_0^{(\$)}$ because, by \Cref{rk: gadgets_matches}.\ref{b_0-b_j_a_i_ortho}, if $ a_{p+1}^\perp \$ \sqcap b_0^{(\$)}b_1\$\neq\emptyset$ (thus $a_{p+1}^\perp \sqcap b_0 b_1\neq\emptyset$) 
    then $\alpha_{p+1}\perp \beta_1$.
    Therefore, under the assumption that $\alpha_{p+i} \not \perp \beta_{i}$ for each $i\in [1,M-p]$, the only possibility, by \Cref{rk: gadgets_matches}.\ref{a_zero_b_i} and \ref{rk: gadgets_matches}.\ref{b_zero_a_i}, would be to align $ a_{p+1}^\perp \$a_0^\perp\$\cdots \$a_0^\perp\$a_M^\perp \$$ with $ b_0^{(\$)}b_1\$ \cdots b_{M-p-1}\$b_0^{(\$)}$.
    This would force the alignment of $a_{even}^\perp\$$ with $b_{M-p}\$$, which is not possible because $M-p$ is odd by Assumption~\ref{ov: M_odd_N_even}. 
    \begin{center}
        \includegraphics{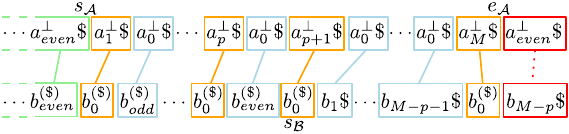}
    \end{center}
    \item[Case $s_{\cB} \le s_\cA$.] The case $e_{\cA} \le e_{\cB}$ has already been considered. If $e_{\cB} < e_{\cA}$, we can proceed  symmetrically to the case $s_{\cA} < s_{\cB}$.
    We briefly sketch the proof. 
    Suppose $\alpha_i \not\perp \beta_N$, for all $i$.
    Then, $b_0^{(\$)}b_N\$$ must align with $a_{q}^\perp\$a_0^\perp\$$, for some $q$. We first prove the fragment $a_{q+1}^\perp\$a_0^\perp\$ a_{q+2}^\perp\$ \cdots\$a_0^\perp\$ a_{M}^\perp\$a_{even}^\perp$ must align with a fragment of $\cB_{suf}$ and by Remarks~\ref{rk: gadgets_matches}.\ref{b_zero_a_i}, \ref{rk: gadgets_matches}.\ref{b_even_odd_not_a_i} and \ref{rk: gadgets_matches}.\ref{a_even_b_even_b_odd}, $q$ must be odd.
    Secondly, assuming that $\alpha_{i}\not \perp \beta_{N-q+i}$ for $i\in [q]$, we get that $a_1^\perp\$a_0^\perp\$ a_{2}^\perp\$ \cdots\$a_0^\perp\$ a_{q}^\perp\$a_0^\perp \$ $ aligns with $b_0^{(\$)}b_{N-q+1}\$\cdots\$ b_0^{(\$)}b_{N}\$ $. This forces the alignment of $a_{even}^\perp\$$ with $b_{N-q}\$$, which is a contradiction since $N-q$ is odd.\qedhere
\end{description}
\end{proof}

\begin{figure}[h!]
        \centering
        \includegraphics[width=.95\textwidth]{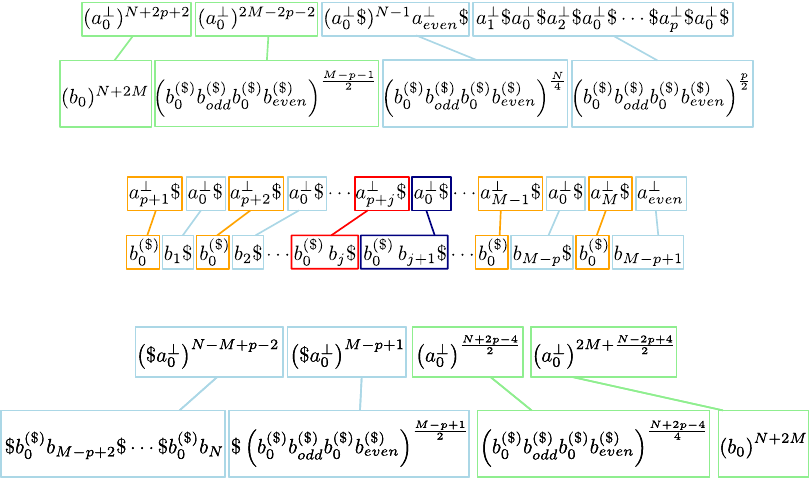}
        \caption{Scheme of an alignment of $\cA$ and $\cB$ in the case $\alpha_i\perp\beta_j$ with $j<i$ and $i\equiv j (\textnormal{mod}~2)$: $p\in[M]$ is such that $p+j=i$ (in particular, $p$ is an even number).}
        \label{fig:intersection_general}
    \end{figure}

\begin{lemma}
    If $\exists~ j\in [N]$ and $i \in [M]$ such that $\alpha_j \perp \beta_i$, then $\cA \sqcap \cB \ne \emptyset$.
\end{lemma}
\begin{proof}
    By Assumption~\ref{ov: ai_bj_i_j_same_parity}, we have $i\equiv j (\textnormal{mod}~2)$.
    Consider the case $i=j$. We show that in this case, $\cA$ and $\cB$ align, which implies that $\cL(\cA)\cap\cL(\cB)\neq\emptyset$. 
    We start by showing how $\cA_{pre}$ aligns with $\cB_{pre}$.
    Using \Cref{rk: gadgets_matches}, we can prove the following facts straightforwardly:
    \begin{itemize}
        \item $(a_0^\perp)^k \sqcap (b_0)^h \ne \emptyset$ for all $k\le h$
        \item $(a_0^\perp)^2 \sqcap b_0^{(\$)}b_{odd}^{(\$)}b_0^{(\$)}b_{even}^{(\$)}\ne \emptyset$
        \item $(a_0^\perp\$)^4\sqcap b_0^{(\$)}b_{odd}^{(\$)}b_0^{(\$)}b_{even}^{(\$)}\ne \emptyset$ and $(a_0^\perp\$)^3a_{even}^\perp\$\sqcap b_0^{(\$)}b_{odd}^{(\$)}b_0^{(\$)}b_{even}^{(\$)}\ne \emptyset$
    \end{itemize}
    Therefore, we can align $\cA_{pre}$ with $\cB_{pre}$ in the following way:
    \begin{center}
    \includegraphics{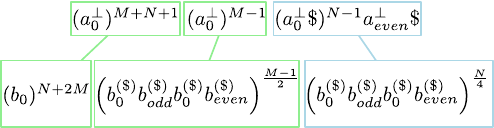}
    \end{center}

    We then show how to align $\cA^\perp$ with $\cB^\perp$.
    By Assumption~\ref{ov: a1_aM_not_orthogonal} we have $i\ne 1$, and by \Cref{rk: gadgets_matches}.\ref{a_zero_b_i} and \ref{rk: gadgets_matches}.\ref{b_zero_a_i} we can align $a_k^\perp\$ a_0^\perp\$ $ with $b_0^{(\$)}b_k\$$ for all $k\in [1,i-1]$. 
    Since $\alpha_i\perp \beta_i$, by \Cref{rk: gadgets_matches}.\ref{b_0-b_j_a_i_ortho} we can align $ a_i^\perp\$  a_0^\perp\$$ with $b_0^{(\$)}b_i\$b_0^{(\$)}b_{i+1}\$$.
    Again by Assumption~\ref{ov: a1_aM_not_orthogonal} we have $i\ne M$, thus we can align $a_k^\perp\$a_0^\perp\$$ with $b_0^{(\$)}b_{k+1}\$ $ for all $k\in [i+1,M-1]$. For $k=M$, we have to align $a_M^\perp\$a_{even}^\perp $ with $b_0^{(\$)}b_{M+1} $, which is possible by \Cref{rk: gadgets_matches}.\ref{a_even_b_p_even} since by Assumption~\ref{ov: M_odd_N_even} $M$ is odd, thus $M+1$ is even.

    \begin{center}
    \includegraphics{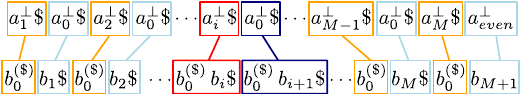}
    \end{center}

    Similarly to the first part of the proof, we can align $\cA_{suf}$ with the remaining part of $\cB$. 
    \begin{center}
    \includegraphics{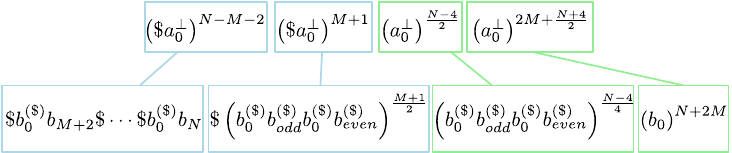}
    \end{center}
    In the general case where $i\equiv j (\textnormal{mod}~2)$, i.e. $|i-j| =p$ for an even $p$ (see Assumption~\ref{ov: ai_bj_i_j_same_parity}), the proof is analogous: 
Figure~\ref{fig:intersection_general} outlines an alignment of $\cB$ with $\cA$ when $j< i$.  
\end{proof}

    Since by construction the sizes of $\cA$ and $\cB$ are $\mO((N+M)d) = \mO(Nd)$, we have obtained Theorem~\ref{thm:hardness}.
    \begin{thm}\label{thm:hardness}
        The $(\conc +,\conc |)$-intersection testing problem cannot be solved in time $\mO((mn)^{1-\alpha})$, for any $\alpha>0$, unless SETH fails.
    \end{thm}

\section{Linear-Time Algorithms}\label{sec:upper_bounds}

This section presents linear-time algorithms for solving the intersection testing problem for all combinations of types of depth-2 homogeneous regular expressions except for those listed in Corollaries~\ref{cor:star-plus} and~\ref{cor: conc_something-conc_star} and Theorem~\ref{thm:hardness}.

\begin{thm}
\label{thm:cdot-or}
    The $(\conc|,t)$-intersection testing problem can be solved in $\mO(m+n)$ time for any type $t\in \{\conc |,|\conc,+\conc\}$.
\end{thm}

\begin{proof}
    Let $\cA$ be a regexp of type $\conc |$ of size $m$, and consider its tree representation. 
    For each node $i$ on the first level, we denote by $\cA_i$ the set of leaves descending from $i$. 
    We can represent $\cA$ as the concatenation of these sets of letters $\cA = \cA_1\conc \cA_2\conc\cdots \conc\cA_k$. 
    Given $\cB$ a regexp of type $t$ and size $n$, we propose different solutions depending on its type $t$.
    \begin{description}
        \item[$t=\conc |$.] We can represent $\cB$ as a concatenation of  sets of letters $\cB = \cB_1\conc \cB_2 \conc \cdots \conc\cB_h$. 
        We have that $\cA \sqcap \cB \ne \emptyset$ if and only if $k=h$ and $\cA_i\cap \cB_i\ne \emptyset$ for all $i\in [k]$, which can be checked in $\mO(m+n)$ total time by radix-sorting $\cA_i$ together with $\cB_i$ for all $i\in[k]$. 
        In particular, if $\cA \sqcap \cB \ne \emptyset$ we can express $\cA \sqcap \cB$ as $(\cA_1 \cap \cB_1) \conc (\cA_2 \cap \cB_2) \conc \cdots \conc (\cA_k \cap \cB_h)$, i.e. we can represent the intersection as a regexp of type $\conc|$ in $\mO(m+n)$ time.
        \item[$t=|\conc$.] In this case, $\cB$ is of the form $[T_1 | T_2|\cdots |T_h]$ for some nonempty strings $T_j$. 
        Remark that if $\cA\sqcap \cB \ne \emptyset$ then there exists $T_i$ such that $|T_i| = k$. 
        Therefore, with a linear preprocessing we can discard the strings in $\cB$ that have length different from $k$. 
        This leads to a set $\cB'$ of strings all of the same length, which is known in the literature as a segment of a Generalised Degenerate (GD) string~\cite{fi20}.
        Since regexps of type $\conc |$ are known as indeterminate strings, which in turn are a special case of GD strings, we can use the algorithm of~\cite{fi20} for GD strings intersection to solve the problem in $\mO(m+n)$ time.
        \item[$t=+\conc $.] In this case, $\cB$ is of the form $[T]^+$ for some nonempty string $T$ of length $n$.
        Remark that if $\cA \sqcap\cB \ne\emptyset$ then $k=\ell n$ for some $1\le \ell\le k$. 
        Therefore, to solve the intersection testing problem, we first check if $k=\ell n$, and then answer a membership query for $T^\ell$ and $\cA$ using the linear-time algorithm proposed in~\cite{focs16}.  \qedhere 
    \end{description}
\end{proof}

\begin{thm}
\label{thm:cdot-plus}
    The $(\conc+,t)$-intersection testing problem can be solved in $\mO(m+n)$ time for any type $t\in \{\conc +,|\conc,+\conc\}$.
\end{thm}

\begin{proof}
    Let $\cA$ be a regexp of type $\conc +$ and size $m$.
    The following transformations preserve the language generated by $\cA$: $c^+c^\ell \equiv c^\ell c^+$ and $(c^+)^\ell \equiv c^{\ell-1}c^+$ for any $c \in \Sigma$ and $\ell\in \mathbb{N}$.
    From now on, we denote by $c^{\ell+}$ the regexp $c^{\ell-1}c^+$.
    Applying the above rule, we can rewrite $\cA=a_1^{\cX_1}\conc \ldots \conc a_k^{\cX_k}$ where $\cX_i\in\{\ell_i, \ell_i+\}$, $\ell_i\in[m]$ and $a_i\neq a_{i+1}$ for all $i\in[k-1]$, $k\leq m$, and we can obviously represent each $a_i^{\cX_i}$ with a pair $(a_i,\cX_i)$.
    For example, the regexp $b^+bc^+c^+a$ of \Cref{fig:tree_example} can be rewritten as $b^{2+}c^{2+}a$.

    Given $\cB$ a regexp of type $t$, we propose different linear-time algorithms depending on $t$.
    \begin{description}
        \item[$t=\conc +$.] Let $\cB$ be of type $\conc +$ and consider its decomposition $\cB = b_1^{\cY_1}\conc \ldots \conc b_h^{\cY_h}$. We have $\cA \sqcap \cB \ne \emptyset$ if and only if $k=h$ and $a_i^{\cX_i}\sqcap b_i^{\cY_i} \ne\emptyset$ for all $i\in[k]$.
        A necessary condition for the  intersection of $a_i^{\cX_i}$ and $b_i^{\cY_i}$ to be nonempty is that $a_i=b_i=c$; furthermore, we can divide the following cases, each of which can be checked in $\mO(1)$ time:
        \begin{enumerate}
            \item\label{case1} $\cX_i=x,\cY_i=y$. Then $c^x \sqcap c^y = c^y$ if and only if $x=y$ ($\emptyset$ otherwise);
            \item\label{case2} $\cX_i=x+,\cY_i=y$. Then $c^{x+} \sqcap c^y = c^y$ if and only if $y\ge x$ ($\emptyset$ otherwise);
            \item\label{case3} $\cX_i=x+,\cY_i=y+$. Then $c^{x+} \sqcap c^{y+} = c^{\max\{x,y\}+}$.
        \end{enumerate}
        
         It then suffices to scan $\cA$ and $\cB$ in parallel and test the intersection of $a_i^{\cX_i}$ with $b_i^{\cY_i}$ by checking Cases~\ref{case1}-\ref{case3} to solve the problem in $\mO(m+n)$ time. If $\cA \sqcap \cB\ne \emptyset$, it holds that $\cA \sqcap \cB = (a_1^{\cX_1} \sqcap b_1^{\cY1}) \conc (a_2^{\cX_2} \sqcap b_2^{\cY_2}) \cdots (a_k^{\cX_k} \sqcap b_k^{\cY_k})$, thus we can express the intersection as a regexp of type $\conc +$.
        \item[$t=|\conc $.] $\cB$ is of the form $[T_1 | T_2|\cdots |T_h]$ for some nonempty strings $T_j$. Decomposing each $T_j$ into maximal runs of single letters and checking Cases~\ref{case1} and~\ref{case2} determines if $T_j \in \cL(\cA)$ in $\mO(|T_j|)$ time, thus solving $(+\conc,|\conc)$-intersection testing  in $\mO(m+n)$ total time.
        \item[$t=+\conc$.] $\cB$ equals $[T]^+$ for some nonempty string $T$ of length $n$.
        We decompose $T$ into maximal single-letter runs to obtain $T = b_1^{\ell_1}\cdots b_h^{\ell_h}$ with  $\ell_i \ge 1$ and  $b_i\ne b_{i+1}$ for $i\in[h-1]$, and consider the decomposition of $\cA=a_1^{\cX_1} \cdots a_k^{\cX_k}$.
        If $h>k$, there are strictly more single-letter runs in any string in $\cL(\cB)$ than in any string in $\cL(\cA)$, which implies that the intersection is empty.
        Therefore, we can assume that $h\le k$.
        Let us first consider the case where $b_1\neq b_h$, that is, $T$ starts and ends with runs of different letters. We compare the decompositions of $\cA$ and $T$ from left to right. Clearly, a necessary condition for the intersection to be nonempty is that $a_1=b_1$ and $a_j=b_{\rho(j)}$ for all $j>1$, where $\rho(j)=((j-1)\mod h) + 1$. 
        In particular, let $b_{\rho(j)}^{\ell_{\rho(j)}} = c^{y}$: if $a_j^{\cX_j}=c^{x+}$, the intersection is nonempty if and only if $x\le y$; otherwise, if $a_j^{\cX_j}=c^{x}$, it is nonempty if and only if $x=y$ (see Cases~\ref{case1}-\ref{case2} above). 
        If the intersection between $a_j^{\cX_j}$ and $b_{\rho(j)}^{\ell_{\rho(j)}}$ is successful, we continue comparing $a_{j+1}^{\cX_{j+1}}$ and $b_{\rho(j+1)}^{\ell_{\rho(j+1)}}$. The last step is to compare $a_k^{\cX_k}$ with $b_h^{\ell_h}$ to decide if $\cA\sqcap\cB\neq\emptyset$ (if $\rho(k)\neq h$ then the intersection is for sure empty).
        Since $k=\mO(n)$ in the worst case, this algorithm takes $\mO(m+n)$ time.

        Now consider the case $b_1=b_h$, i.e. $T$ starts and ends with the same letter. The algorithm is the same as the previous case, except every time we need to check the intersection of $b_h^{\ell_h}$ with $a_j^{\cX_j}$ for some $j<k$, we instead check the intersection of $a_j^{\cX_j}$ with $b_h^{\ell_h}b_1^{\ell_1}=b_h^{\ell_h+\ell_1}$, and move on to checking the intersection of $a_{j+1}^{\cX_{j+1}}$ with $b_2^{\ell_2}$ at the next step. This modification does not change the complexity of the algorithm, which still requires $\mO(m+n)$ time. \qedhere
    \end{description}
\end{proof}

\begin{thm}
\label{thm:or-dot_plus-cot}
    The $(t,u)$-intersection testing problem can be solved in $\mO(m+n)$ time for all types $t,u\in \{|\conc,+\conc\}$.
\end{thm}

\begin{proof}
    \begin{description}
        \item[$(|\conc,|\conc)$.] Both $\cA$ and $\cB$ are finite dictionaries of nonempty strings: $\cA=[S_1 | S_2|\cdots |S_k]$, $\cB=[T_1 | T_2|\cdots |T_h]$. Therefore, testing if $\cA\sqcap\cB\neq\emptyset$ is equivalent to checking whether $S_i=T_j$ for some $1\le i\le k,1\le j\le h$. This can be easily done in $O(m+n)$ time by radix-sorting the strings of $\cA$ and $\cB$ all together, then comparing pairs of consecutive strings in the sorted sequence. 
        Since, after each string comparison, one of the two elements is discarded (unless the two strings are equal, in which case we conclude that $\cA \sqcap \cB \neq \emptyset$), comparing the strings letter by letter gives $O(m+n)$ total time.
        \item[$(+\conc,|\conc)$.] Let $\cA$ be a regexp of type $+\conc$ and $\cB$ of type $|\conc $.
        Then, $\cA$ is of the form $[T]^+$ and $\cB$ is of the form $[T_1 | T_2|\cdots |T_h]$ for some nonempty strings $T_j$.
        Since $T_j \in \cL(\cA)$ if and only if $T_j=T^k$ for some $k\geq 1$, and this condition can be verified in  $\mO(|T_j|)$ time with a linear scan of $T_j$, computing the intersection requires $\mO(m+n)$ total time.
        \item[$(+\conc,+\conc)$.] We have $\cA=[T_1]^+$ and $\cB=[T_2]^+$ for some nonempty strings $T_1$, $T_2$.
        We say that a nonempty string $D$ is a divisor of $T$ if $T = D^i$ for some $i\ge 1$.
        A sufficient condition for $\cA \sqcap \cB \ne \emptyset$ is that $T_1 = D^i$ and $T_2=D^j$ for the same nonempty string $D$ (i.e. $D$ is a common divisor), and in particular, $[D]^{\lcm(i,j)+}\subseteq \cA \sqcap \cB$.
        A necessary condition for $\cA \sqcap \cB \ne \emptyset$ is that  $i,j$ exist such that $T_1^i=T_2^j$; using Theorem~1 in \cite{10.1145/990518.990520}, we can deduce that if this is the case, $T_1$ and $T_2$ have a common divisor.
        Therefore, to decide whether the intersection is not empty, it is sufficient and necessary to check if $T_1$ and $T_2$ have a common divisor. 
        This can be done in $\mO(m+n)$ time using Theorem~2 in \cite{10.1145/990518.990520}. \qedhere
    \end{description}
\end{proof}

\begin{thm}
\label{thm:star-cdot}
    The $(*\conc,t)$-intersection testing problem can be solved in $\mO(m+n)$ time for all types $t\in \{\conc|,\conc+,|\conc,+\conc,*\conc\}$.
\end{thm}

\begin{proof}
    Let $\cA = [T]^*$ of type $*\conc$, and define $\cA' = [T]^+$ of type $+\conc$.
    For all regular expressions $\cB$ of type $t\in \{\conc|,\conc+,|\conc,+\conc\}$, we have that $\epsilon\notin \cL(\cB)$, therefore $\cA \sqcap \cB = \cA' \sqcap \cB$.
    Thus we can reduce $(*\conc,t)$-intersection to $(+\conc,t)$-intersection for all types $t\in \{\conc|,\conc+,|\conc,+\conc\}$ and obtain the statement from Theorems~\ref{thm:cdot-or},~\ref{thm:cdot-plus} and~\ref{thm:or-dot_plus-cot}.
    The remaining case of $(*\conc,*\conc)$-intersection is trivial since $\epsilon \in \cA \sqcap \cB$ for all $\cA,\cB$ of type $*\conc$. Furthermore, to compute explicitly the whole intersection, one can consider $\cA'$, $\cB'$ obtained for $\cA$, $\cB$ by replacing the root label $*$ with $+$, and then apply the algorithm for $(+\conc,+\conc)$-intersection described in Theorem~\ref{thm:or-dot_plus-cot}. 
\end{proof}

Intersection testing for the remaining types of regular expressions is always solvable in linear time, independently of the type of the other regexp, as proved in Theorems~\ref{thm:star_plus-or} and~\ref{thm:or-star_plus}.

\begin{thm}
\label{thm:star_plus-or}
    The $(+|,t)$-intersection testing problem and the $(*|,t)$-intersection testing problem can be solved in $\mO(m+n)$ time for any type $t$ of homogeneous regexp of depth 2.
\end{thm}
\begin{proof}
Let $\cA = [c_1|\cdots | c_m]^+$ be of type $+|$. In this case we have $\cL(\cA)=\realph{\cA}^+$, and analogously, when $\cA$ is of type $*|$, its language is $\realph{\cA}^*$. 
To solve intersection testing for $\cA$ of type $+|$ or $*|$ and any other homogeneous depth-$2$ regular expression $\cB$, it thus suffices to check whether $\cL(\cB)$ contains a string over the alphabet $\realph{\cA}$ (including the empty string if $\cA$ is of type $*|$). 
Given a letter $c$ in $ \realph{\cA}\cup \realph{\cB}$,
we can check in constant time whether $c$ belongs to $\realph{\cA}$ by radix-sorting $\realph{\cA}$ and $\realph{\cB}$ together, keeping track of which letters belong to either set and removing duplicates, and storing 
a map for the rank of $c$ in $ \realph{\cA}\cup \realph{\cB}$. 
This provides an $\mO(m+n)$-time solution for all types listed below (the types $t\in\{*+,+*\}$ were already considered in Corollary~\ref{cor:star-plus}), as a linear scan of $\cB$ and/or of the sorted sequence $ \realph{\cA}\cup \realph{\cB}$ then suffices to solve the problem.
We provide more details for the case where $\cA$ is of type $+|$: the only difference when $\cA$ is of type $*|$ is having to consider also the empty string $\epsilon$.
\begin{description}
\item[$t\in\{+|, *|\}$.] It suffices to scan the sorted sequence to check whether $\realph{\cA} \cap \realph{\cB } \neq \emptyset$.

\item[$t \in\{\conc +,  + \conc, * \conc\}$.] The intersection is nonempty if and only if $\realph{\cB}\subseteq  \realph{\cA}$, which again can be checked by scanning the sorted sequence once.

\item[$t =\conc *$.] The intersection is nonempty if all letters occurring in $\cB$ without a $*$ (i.e. direct children of the root in the tree representation) belong to $\realph{\cA}$.

\item [$t \in\{|*, |+\}$.] Here, $\cB$ is a set of single-letter runs, it thus suffices to check whether the letter of one of such runs is in $\realph{\cA}$.

\item[$t\in\{\circ |, | \circ \}$.] In the first case, $\cB$ is a concatenation of finite sets of letters, and it suffices to check whether in each set there is at least one letter that belongs to $\cA$. In the second case, $\cB$ is a finite dictionary of nonempty strings, and it suffices to check whether the alphabet of at least one of them is contained in $\realph{\cA}$. \qedhere
\end{description}
\end{proof}

\begin{thm}
\label{thm:or-star_plus}
    The $(|+,t)$-intersection testing problem and the $(|*,t)$-intersection testing problem can be solved in $\mO(m+n)$ time for any type $t$ of homogeneous regexp of depth 2.
\end{thm}
\begin{proof}
Let $\cA$ be of type $|+$. 
Then, $\cA$ is of the form $[T_1|\cdots |T_m]$, with $T_j = c_j$ or $T_j = c_j^+$ for some $c_j \in \Sigma$ (i.e. $T_j$ is either a single letter or a single-letter run).
Let $\realph{\cA} = \{c_1,...,c_k\}$ be the set of letters used by $\cA$. 
Let us consider all possible types $t$ for $\cB$.
The cases $t\in \{+*,*+,+|,*|\}$ were considered in Corollary~\ref{cor:star-plus} and Theorem~\ref{thm:star_plus-or}.
Solutions for the other types follow.
 \begin{description}
    \item[$t\in\{|+,|*\}$.] $\cB$ is a disjunction of runs, it thus suffices to check whether $\realph{\cA} \cap \realph{\cB } \neq \emptyset$, proceeding as in Theorem~\ref{thm:star_plus-or}.
    
    \item[$t=|\conc$.] $\cB$ is a dictionary $[S_1|\cdots | S_h]$, where each $S_i$ is a nonempty string. We can proceed as in case $(|\conc,|\conc)$ of Theorem~\ref{thm:or-dot_plus-cot}.
    
    \item[$t=\conc |$.] $\cB$ is of the form  $\cB_1\conc \cB_2\conc\cdots \conc\cB_h$, where each $\cB_i$ is a set of letters. Let $\realph{\cA}^{(+)}\subseteq\realph{\cA}$ be the set of letters that occur in $\cA$ with a $+$: the intersection is nonempty if and only if $\cL(\cB)$ contains a run of length $h$ of some letter in $\realph{\cA}^{(+)}$. 
    To check this condition, we scan $\cB$ maintaining a counter for each letter in $\realph{\cB}$ to find the subset $\realph{\cB}^{(h)}$ of letters that occur $h$ times, corresponding to all single-letter runs of $\cL(\cB)$. If $\realph{\cB}^{(h)}$ is nonempty, we radix-sort it with $\realph{\cA}^{(+)}$ and scan the sorted sequence to find the answer in $\mO(m+n)$ total time.
    
    \item[$t=\conc *$.] $\cB$ is a concatenation of symbols
    $b_i$ or $b_i^*$ with  $b_i\in\realph{\cB}$. Then, $\cA \sqcap \cB \ne \emptyset$ if and only if either (i) there is only one letter $b\in\realph{\cB}$ occurring in $\cB$ without $*$, and $b\in\realph{\cA}$;
    or (ii) no letter occurs in $\cB$ without $*$, and  $\realph{\cA}\cap \realph{\cB}\neq\emptyset$. Both of these conditions can be trivially checked in $\mO(m+n)$ time. 
    
    \item[$t=\conc +$.] $\cB$ can be made into the form  $b_1^{\cY_1}\cdots b_h^{\cY_h}$ using the transformation in Theorem~\ref{thm:cdot-plus}, where $\cY_i \in \{\ell_i,\ell_i+\}$, $\ell_i\in[m]$ and $b_i \ne b_{i+1}$ for all $i\in [h-1]$. Then, $\cA \sqcap \cB \ne \emptyset$ if and only if $h=1$ and $b_1\in\realph{\cA}$.
    
    \item[$t\in\{+\conc,*\conc\}$.] $\cB$ is of the form $[T]^+$, (resp. $[T]^*$) for some nonempty string $T$. For the intersection to be nonempty,  we need that $T=c^\ell$ for some $c \in \Sigma_{\cA}$, with the further constraint $\ell=1$ if $c$ appears in $\cA$ without the operator $+$ (resp. $*$). This can be trivially checked in $\mO(m+n)$ time. 
\end{description}
Analogous solutions work for $\cA$ of type $|*$, the only difference being having to consider the possibility that $\epsilon\in\cL(\cA)$.
\end{proof}

\bibliography{references}

\end{document}